\documentclass[preprint,prx,amsmath,amssymb,longbibliography,superscriptaddress]{revtex4-1}

\pdfoutput=1

\usepackage{graphicx}
\usepackage{amsmath,latexsym,tabularx}
\usepackage{xcolor}
\usepackage{multirow}
\usepackage[
  colorlinks=true,
  urlcolor=blue,
  linkcolor=blue,
  citecolor=blue
]{hyperref}
\usepackage{multirow,booktabs}
\usepackage{float}
\usepackage{upgreek}

\usepackage{setspace}
\usepackage{caption}

\newcommand{\affilLL}[0]{MIT Lincoln Laboratory, Lexington, Massachusetts 02421, USA}
\newcommand{\affilMIT}{Massachusetts Institute of Technology, Cambridge, Massachusetts 02139, USA}

\newcommand{\unit}[1]{\, \mathrm{#1}}
\newcommand{\sr}{Sr$^{+}$}

\newcommand{\specificthanks}[1]{\@fnsymbol{#1}}
\newcommand{\mus}{$\upmu$s}

\begin{document}

\title{\normalsize Optical Atomic Clock Interrogation Via an Integrated Spiral Cavity Laser}


\author{William Loh}
\thanks{\setstretch{0.9} These authors contributed equally to this work.\newline \vspace{-2em} Correspondence to william.loh@ll.mit.edu, david.reens@ll.mit.edu, and robert.mcconnell@ll.mit.edu.}
\affiliation{\affilLL}

\author{David Reens}
\thanks{\setstretch{0.9} These authors contributed equally to this work.\newline \vspace{-2em} Correspondence to william.loh@ll.mit.edu, david.reens@ll.mit.edu, and robert.mcconnell@ll.mit.edu.}
\affiliation{\affilLL}

\author{Dave Kharas}
\affiliation{\affilLL}

\author{Alkesh Sumant}
\affiliation{\affilLL}

\author{Connor Belanger}
\affiliation{\affilLL}

\author{Ryan T. Maxson}
\affiliation{\affilLL}

\author{Alexander Medeiros}
\affiliation{\affilLL}

\author{William Setzer}
\affiliation{\affilLL}

\author{Dodd Gray}
\affiliation{\affilLL}

\author{Kyle DeBry}
\affiliation{\affilLL}
\affiliation{\affilMIT}

\author{Colin D. Bruzewicz}
\affiliation{\affilLL}

\author{Jason Plant}
\affiliation{\affilLL}

\author{John Liddell}
\affiliation{\affilLL}

\author{Gavin N. West}
\affiliation{\affilMIT}

\author{Sagar Doshi}
\affiliation{\affilMIT}

\author{Matthew Roychowdhury}
\affiliation{\affilLL}

\author{May Kim}
\affiliation{\affilLL}

\author{Danielle Braje}
\affiliation{\affilLL}

\author{Paul W. Juodawlkis}
\affiliation{\affilLL}

\author{John Chiaverini}
\affiliation{\affilLL}
\affiliation{\affilMIT}

\author{Robert McConnell}
\affiliation{\affilLL}

\begin{abstract}

Optical atomic clocks have demonstrated revolutionary advances in precision timekeeping \cite{Hinkley2013, Bloom2014, Brewer2019, Koller2017, Godun2014, Huntemann2016}, but their applicability to the real world is critically dependent on whether such clocks can operate outside a laboratory setting \cite{Leibrandt2011, Koller2017, Liu2018, Grotti2018, Takamoto2020, Stuhler2021, Cao2022}. The challenge to clock portability stems from the many obstacles
not only in miniaturizing the underlying components of the clock---namely the ultrastable laser, the frequency comb, and the atomic reference itself---but also in making the clock resilient to environmental fluctuations. Photonic integration offers one compelling solution to simultaneously address the problems of miniaturization and ruggedization \cite{Mehta2020, Niffenegger2020}, but brings with it a new set of challenges in recreating the functionality of an optical clock using chip-scale building blocks. The clock laser used for atom interrogation is one particular point of uncertainty, as the performance of the meticulously-engineered bulk-cavity stabilized lasers \cite{Young1999, Jiang2011, Kessler2012} would be exceptionally difficult to transfer to chip. Here we demonstrate that a chip-integrated ultrahigh quality factor (Q) spiral cavity, when interfaced with a 1348 nm seed laser, reaches a fractional frequency instability of $7.5 \times 10^{-14}$, meeting the stability requirements for interrogating the narrow-linewidth transition of $^{88}$Sr$^+$ upon frequency doubling to 674 nm. In addition to achieving the record for laser stability on chip, we use this laser to showcase the operation of a Sr-ion clock with short-term instability averaging down as $3.9 \times 10^{-14} / \sqrt{\tau}$, where $\tau$ is the averaging time. Our demonstration of an optical atomic clock interrogated by an integrated spiral cavity laser opens the door for future advanced clock systems to be entirely constructed using lightweight, portable, and mass-manufacturable integrated optics and electronics.

\end{abstract}

\maketitle


\section{Main Text}
Optical atomic clocks offer unprecedented levels of precision in measurements of frequency, time, and position by tapping into the inherent stability afforded by locking an ultrahigh-frequency laser to a reference atom's narrow-linewidth electronic transitions \cite{Ludlow2015}. Over the last two decades, optical clocks have improved to the point of outperforming the best microwave clocks by two orders of magnitude \cite{Brewer2019}. This increased precision is not only important for improving the performance of future communications and navigation networks, but also for enabling new and emerging science, such as geodetic measurements of the earth, searches for dark matter, and investigations into possible long-term variations of fundamental physics constants \cite{Maleki2005, Ludlow2015, McGrew2018}. 
In recent years, much effort has been devoted to miniaturizing optical clocks, driven by the necessity for bringing them into the field so that they may be used in real-world applications. Such efforts have paved the way for transportable optical clocks able to operate outside the confines of a laboratory \cite{Koller2017, Takamoto2020}. 
While additional reductions in size, weight, and power can be achieved from further scaling down the numerous individual pieces that comprise an optical clock, extant transportable narrow-linewidth optical clocks remain at the m$^3$ scale. Integrated photonics offers the possibility of dramatically miniaturizing sophisticated system functionality to the size scale of a semiconductor chip, including the light delivery and collection processes vital to atom interrogation in an optical clock. Beyond a small form factor, photonic integration also provides the means for inherent rigidity and path stability~\cite {Niffenegger2020}, albeit at the cost of necessitating a complete redesign of the clock using chip-scale components.


The ultrastable laser represents one key impediment to clock integration, as current clock lasers are all referenced to large and unwieldy Fabry-Perot cavities operating under conditions of ultrahigh vacuum and multiple layers of temperature shielding and stabilization, with the most advanced laser cavities requiring cryogenic temperatures \cite{Kessler2012, Robinson2019}. The large cavity is an intended feature of such lasers, since the stability of the cavity resonance frequency against thermal perturbations improves with length and mirror spot size \cite {Numata2004, T.Kessler2012}. Despite these benefits of cavity volume, a viable path towards miniaturizing such Fabry-Perot cavities has been shown \cite {Davila-Rodriguez2017, Didier2019, Jin2022}, trading off only partly in performance and still achieving thermal-noise limited stability at levels better than $5 \times 10^{-14}$ with modest levels of vacuum \cite {Liu2023}. However, as optical reference cavities are brought on chip to enable further size reduction and future integration with the rest of the clock components, the combination of large and unavoidable thermal noise fluctuations coupled with a temperature-dependent refractive index results in significantly degraded cavity stability \cite{Gorodetsky2004, Matsko2007, Huang2019, Panuski2020}. Recently, considerable effort has been dedicated to narrowing the linewidth of integrated lasers \cite {Lee2012, Loh2015, Gundavarapu2019, Xiang2023}, with the most successful demonstrations being those that employ long cavity lengths as a method of averaging the thermal fluctuations in one cavity's round trip. Cavity lengths on the order of meters can be achieved in small volumes by wrapping a waveguide around the circumference of a chip hundreds of times to form a spiral resonator, exploiting the ability to place waveguides in close proximity to one another with negligible crosstalk~\cite {Lee2013, Guo2022, Liu2022}. To date, the best stability reported from an integrated resonator is $1.8 \times 10^{-13}$ for a linewidth of 36 Hz at 1550 nm after subtraction of linear drift \cite {Liu2022}. However, due to the stringent stability requirements imposed on the clock laser, it is as of yet still uncertain whether integrated lasers are capable of producing a linewidth narrow enough to lock to an atom such as to outperform extant vapor cells and microwave atomic clocks.

Here, we showcase a seed laser stabilized to an integrated spiral cavity that achieves a fractional frequency instability of $7.5 \times 10^{-14}$ at 1348 nm, or a linewidth of 16.7 Hz, without making use of any drift subtraction. This integrated spiral cavity laser (ISCL) not only improves on previous best-in-class demonstrations of chip-integrated reference cavities by a factor of 2.4, but also operates at a wavelength that is---upon frequency doubling---coincident with the 0.4 Hz linewidth 5$S_{1/2} \longleftrightarrow 4D_{5/2}$ quadrupole transition in the commonly-used clock ion $^{88}$Sr$^+$ \cite{Barwood1999, Dube2013, nichol2024network}. To demonstrate the promise of an integrated approach to an optical clock, we send the frequency doubled output of our ISCL to a single trapped strontium ion and show the ability to lock the laser's frequency to the ion's narrow-linewidth transition at 674 nm. One key challenge to our clock implementation is the laser's frequency drift, which ubiquitously degrades the stability of all on-chip light sources and occurs at a similar time scale to that used in the feedback loop for locking the laser to the atom. 
We overcome this difficulty by developing advanced protocols for spectroscopy and locking to the atomic reference, and we use these protocols to perform a clock self-comparison by interleaving two independent clock locking sequences on the ion. The short-term fractional frequency instability achieved closely follows that of a recent demonstration of a Sr-ion clock operated with a fiber stimulated Brillouin scattering (SBS) laser in which significant effort was devoted to cancelling the laser drift \cite{Loh2019, Loh2020}, and is an order of magnitude more stable than state-of-the-art microwave clocks \cite{Jefferts2000}.


\begin{figure}[t b !]
\includegraphics[width = 0.95 \columnwidth]{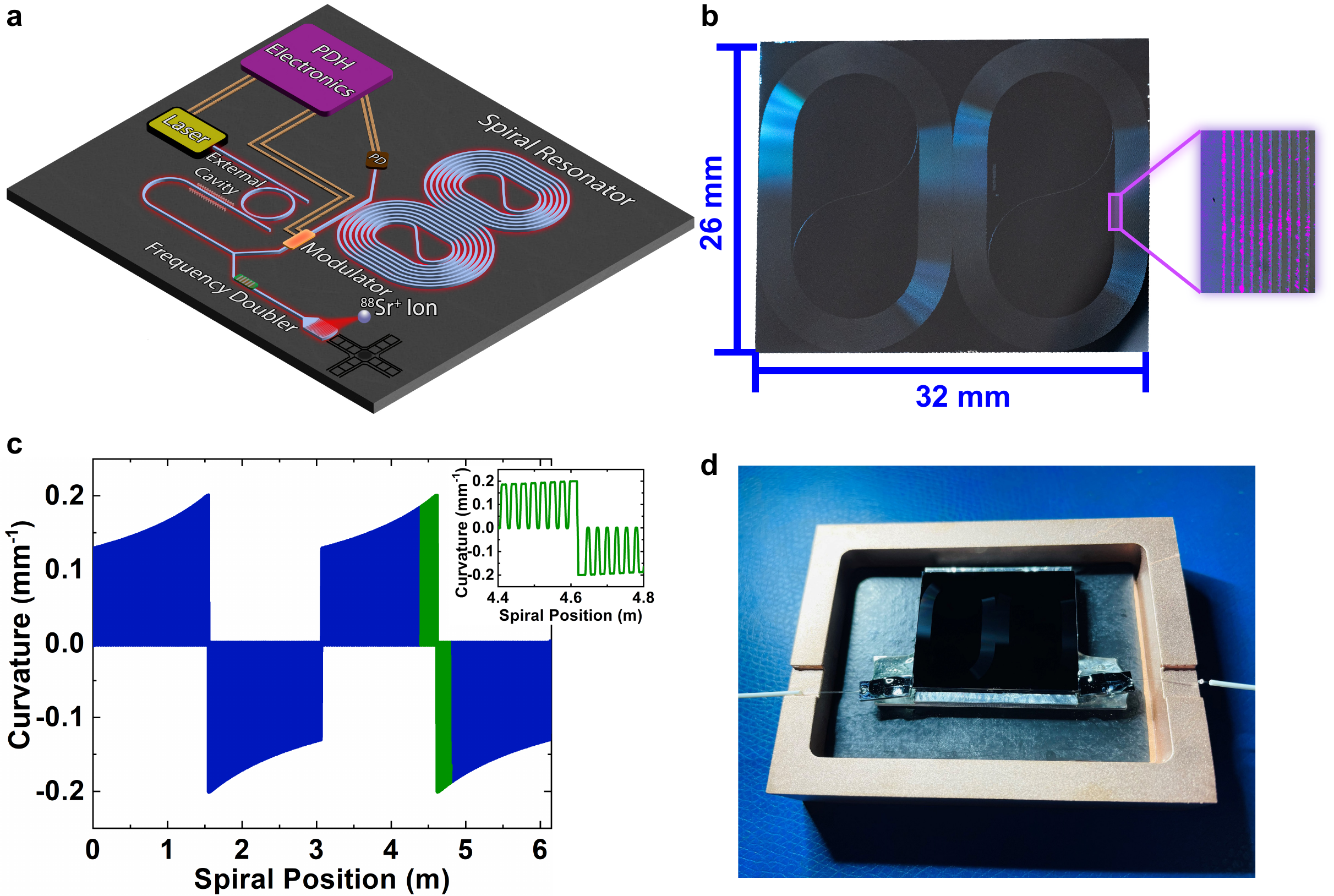}
\caption{
    \textbf{ISCL laser design.}
    \textbf{a}, Illustration of an eventual fully integrated clock laser comprising an external cavity seed laser probing an ultrahigh quality factor spiral resonator, a phase modulator and photodetector to enable Pound-Drever-Hall (PDH) \cite{Drever1983} locking of the seed to the resonator, a frequency doubler for reaching 674 nm, and a trapped strontium ion for serving as the frequency reference.
    \textbf{b}, Photograph of the spiral resonator chip consisting of 6.1 meters of path length. A zoomed in section of the resonator shows a series of waveguides 15 $\upmu$m in width and spaced by 40 $\upmu$m, overlapped with an infrared image of the circulating light.
    \textbf{c}, Spiral resonator design showing the curvature of the waveguide along the propagation path for light. The inset shows a zoomed in image of the curvature at the transition between clockwise and counterclockwise spiraling, corresponding to positions between 4.4 m and 4.8 m along the spiral path (green).
    \textbf{d}, Photograph of the fiber attached spiral resonator coupled at both the input and output facets of the chip. The chip is mounted in a copper enclosure for thermal stabilization.
}
\label{fig:fig1}
\end{figure}

The operation of our clock laser relies on the interaction between a number of components necessary to accomplish locking of a seed laser to a high-Q spiral resonator and the subsequent frequency doubling of the stabilized output to reach 674 nm, all of which are compatible with chip integration. Figure \ref{fig:fig1}a illustrates our vision for a fully chip-integrated clock laser comprising the following components: an external cavity laser and photodetector---both operational at 1348 nm and implemented using InGaAsP photonics, an ultralow-loss passive spiral resonator formed through Si$_3$N$_4$/SiO$_2$, and a phase modulator and frequency doubler created in LiNbO$_3$. In this work, we primarily investigate the performance of the integrated spiral resonator, as its stability is paramount for determining the ultimate stability of the light at 674 nm. The other ancillary components in the clock laser are all devices that are currently packaged separately, but that could be integrated onto a single monolithic chip in the future.

Figure \ref{fig:fig1}b shows a photograph of the spiral resonator. We design the resonator with the intent to maximize the path length within a 26 mm × 32 mm rectangle constrained by the size of a fabrication reticle. The resonator consists of two spirals of waveguide, conjoined at the center of the chip, with each spiral consisting of a segment of waveguide coiling inward followed by a turn at the center and then a segment of waveguide coiling outward. Our chosen spiral resonator design comprises a total length of 6.1 meters, a minimum bend radius of 5 mm, a waveguide spacing of 40 $\upmu$m, and a waveguide width of 15 $\upmu$m. Figure \ref{fig:fig1}c shows the curvature of the resonator plotted along the waveguide path with the sign of the curvature alternating according to the direction of the spiraling as seen by the light propagating along the path. In all cases, the curvature remains below 0.2 mm$^{-1}$ in magnitude. A bus waveguide comes into close proximity to one of the spiral loops to permit coupling into and out of the resonator. We couple into the bus waveguide via fibers attached on both the input and output sides (Fig.~\ref{fig:fig1}d) and place the resonator in a temperature controlled copper enclosure.


\begin{figure}[t b !]
\includegraphics[width = 0.95 \columnwidth]{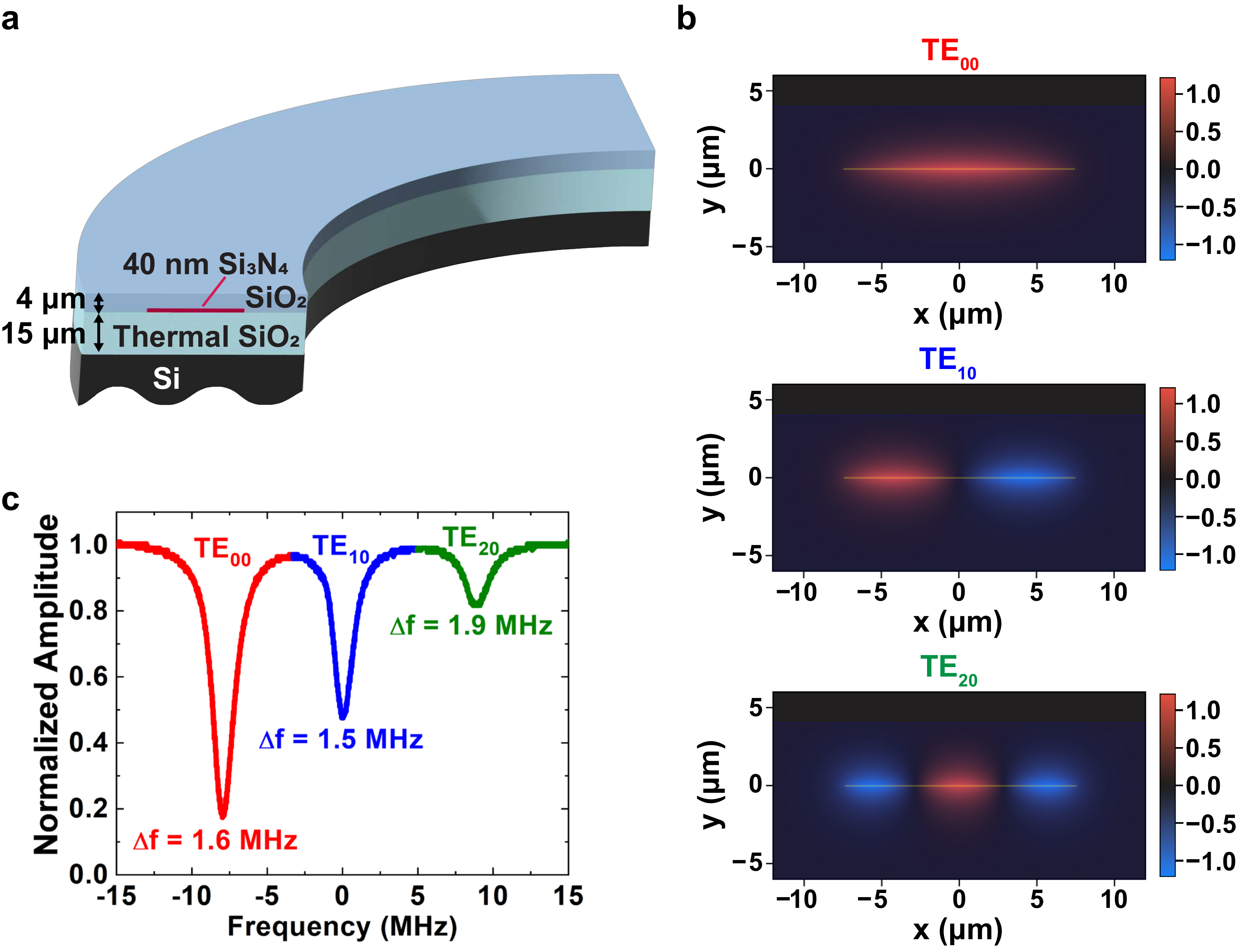}
\caption{
    \textbf{Spiral waveguide modes.}
    \textbf{a}, Spiral waveguide device layers comprising a Si$_3$N$_4$ core and SiO$_2$ cladding. The combination of layers forms a dilute mode with ultralow loss.
    \textbf{b}, Spiral waveguide transverse modes supported by the device structure. The modes expand outside the nitride core to a large degree, and primarily reside in the oxide cladding.
    \textbf{c}, Laser scan over the three transverse modes of the spiral resonator. The modes are dispersed in frequency over a span of $\sim$20 MHz and are able to be individually probed. The TE$_{00}$, TE$_{10}$, and TE$_{20}$ modes exhibit linewidths of 1.6 MHz, 1.5 MHz, and 1.9 MHz, respectively, and have varied levels of mode extinction.
}
\label{fig:fig2}
\end{figure}

A cross-section of the spiral resonator is illustrated in Fig.~\ref{fig:fig2}a. The structure consists of a thin 40 nm Si$_3$N$_4$ waveguide core surrounded by low-loss SiO$_2$ as the cladding \cite {Bauters2011}. A total of 15 $\upmu$m of thermal oxide resides below the nitride waveguide, resting on a silicon substrate, while above the core is 4 $\upmu$m of tetraethyl orthosilicate (TEOS) oxide deposited using low pressure chemical vapor deposition (LPCVD). The combination of a thin waveguide core and thick cladding creates a dilute optical mode that minimizes the scattering loss at the sidewalls of the nitride and also the absorption loss within the nitride itself. The use of a 15 $\upmu$m wide nitride core further decreases the overlap, and thus power loss, of the optical mode to the waveguide sidewalls. However, because the core width is several times larger than the optical wavelength it supports, multiple higher order transverse modes are confined by the waveguide geometry. These transverse modes are visualized in Fig.~\ref{fig:fig2}b, which includes the fundamental TE$_{00}$ mode along with the first and second higher order TE$_{10}$ and TE$_{20}$ modes, having simulated mode areas of 27 $\upmu$m$^2$, 29 $\upmu$m$^2$, and 31 $\upmu$m$^2$, respectively. With a resonant structure, as in the case of the spiral, these modes become spectrally resolvable.

A measurement of the seed laser scanned across the spiral resonator's three transverse modes (Fig.~\ref{fig:fig2}c) shows the individual modes dispersed in frequency across a span of $\sim$20 MHz. This pattern of modes is repeated every 33.7 MHz, corresponding to the spiral resonator's free spectral range. The fundamental mode, having the highest group index, is located on the lower frequency side of the scan, and exhibits a linewidth of 1.6 MHz corresponding to an intrinsic Q of $2.0 \times 10^8$. For the results presented in this work, we stabilized the seed laser to the TE$_{10}$ mode. This lock point offers a comparable Q to the fundamental mode, but exhibits a lower mode extinction of 52$\%$ than the fundamental. This allows more optical power to be diverted past the resonator for Pound-Drever-Hall error detection, thereby minimizing power-dependent frequency shifts of the optical resonance.


\begin{figure}[t b !]
\includegraphics[width = 0.95 \columnwidth]{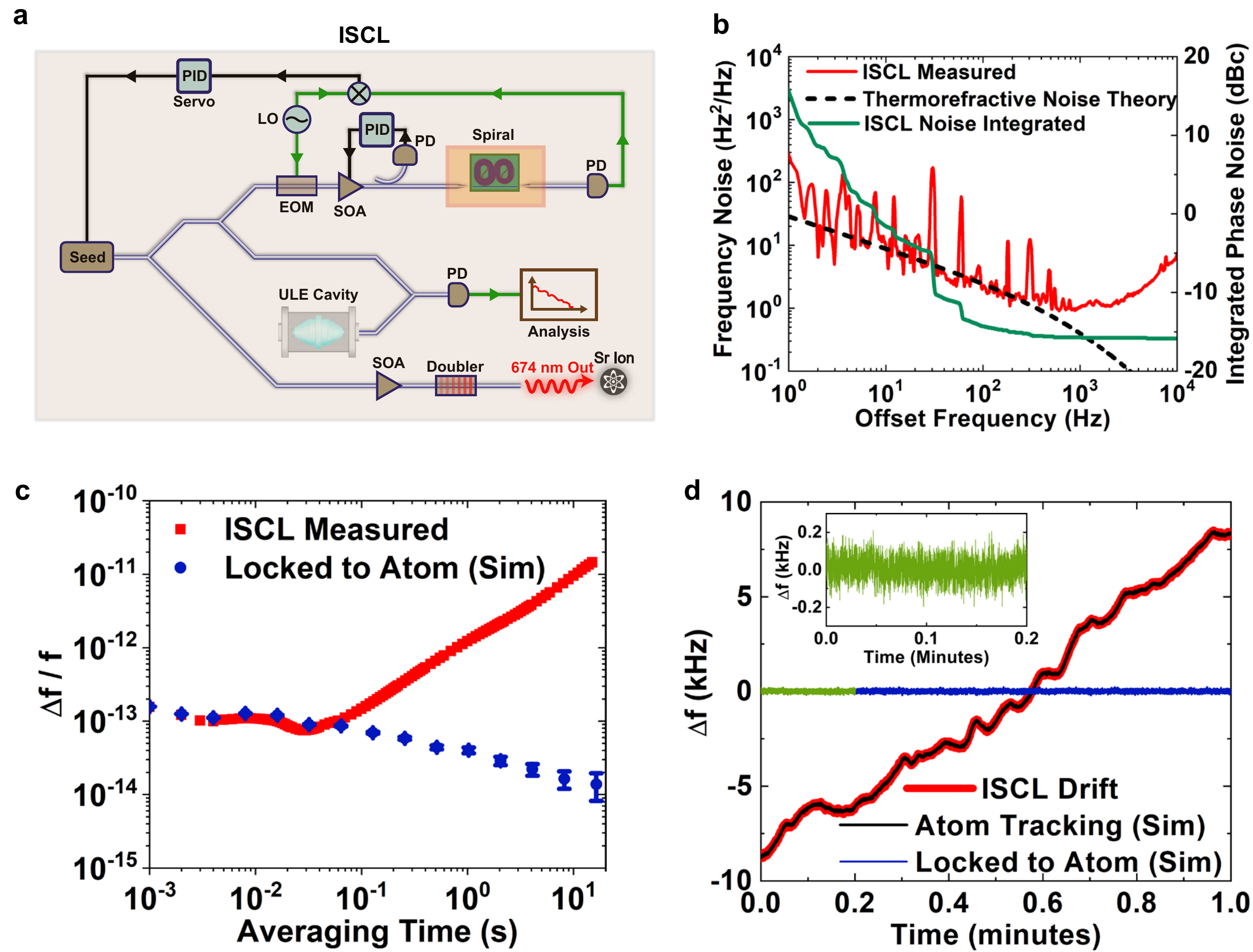}
\caption{
    \textbf{ISCL Measurements.}
    \textbf{a}, ISCL system diagram comprising a seed laser that probes the resonance of the spiral reference cavity. Various components are necessary to enable the generation of ultra-narrow linewidth light at 1348 nm and the conversion to 674 nm for use with the strontium ion.
    \textbf{b}, Frequency noise measurement of the ISCL (red solid line). The calculated thermorefractive noise limit (black dashed line) is shown for comparison. A cumulative integration (green solid line) over the ISCL's phase noise from high-to-low offset frequencies highlights the contribution of various sections of the noise spectral density to the total laser noise.
    \textbf{c}, Allan deviation measured for the ISCL (red solid squares) indicating a minimum fractional frequency instability of $7.5 \times 10^{-14}$ at 30 ms. A simulation of the Sr-ion clock performance using the ISCL as the basis for interrogation (blue solid circles) indicates a $\Delta f / f$ of $4.0 \times 10^{-14}$ at 1 s. 
    \textbf{d}, Measured time series of the ISCL frequency (red solid line) over a one minute span with an overall drift rate of 280 Hz/s. A simulation of the applied laser frequency corrections to track the atom transition frequency (black solid line) is also shown. Numerically applying these corrections to the measured laser drift (blue solid line) indicates stable locking to the atom. The inset shows 0.2 minutes of the stabilized laser (green) exhibiting fluctuations of $\pm$100 Hz.
}
\label{fig:fig3}
\end{figure}

The configuration of the ISCL is depicted in Fig.~\ref{fig:fig3}a and consists of a semiconductor external cavity seed laser operating at 1348 nm that has its light split three ways. A quarter of its 20 mW output power is diverted to probing the spiral resonator. This light is phase modulated by an electro-optic modulator (EOM) and amplified by a semiconductor optical amplifier (SOA) before reaching the spiral resonator. The output of the resonator is sent to a photodetector (PD) for demodulation of the PDH error signal and finally to a proportional-integral-derivative (PID) servo that maintains the seed laser frequency on resonance. A 10$\%$ tap after the SOA is also used in a servo loop to keep the power into the spiral resonator constant against fluctuations. Half of the locked seed laser light is sent to a second SOA and a frequency doubler to generate the ultra-narrow linewidth 674 nm light for probing the strontium ion. The remaining 25$\%$ of the locked seed laser power is used to heterodyne against a 1348 nm ultralow-expansion (ULE) cavity stabilized laser that serves solely as a reference for analyzing the performance of the ISCL.

Figure \ref{fig:fig3}b presents the measured frequency noise of the ISCL at low offset frequencies between 1 Hz and 10 kHz most relevant to atom interrogation. The measured noise closely follows the calculated thermorefractive noise limit for a resonator having a mode area of 29 $\upmu$m$^2$ and a length of 6.1 m. A cumulative integral of the ISCL's phase noise from 1 MHz down towards lower frequencies indicates significant noise contributions arising from the power line at frequencies of 60 Hz, 180 Hz, and 300 Hz and from building fans operating at 30 Hz. Above 10 kHz offset frequency, the servo gain of the spiral resonator lock and the ULE cavity lock both degrade to zero, and the measured noise exhibits a servo bump at 350 kHz that is a combination of both the ISCL and the ULE reference. The integrated contribution of this noise is relatively small and negligibly impacts the performance of the optical clock.

The measured fractional frequency instability ($\Delta f / f$) of the ISCL (Fig.~\ref{fig:fig3}c) reaches a minimum of $7.5 \times 10^{-14}$ at a timescale of 30 ms, corresponding to a linewidth of 16.7 Hz for 1348 nm light. Throughout the range of averaging times between 3 ms and 60 ms, $\Delta f / f$ remains at or below $1.0 \times 10^{-13}$. At longer timescales, the frequency drift exhibited by the ISCL is expected to be compensated for by the atomic reference. 

A time series measurement of the ISCL's frequency (Fig.~\ref{fig:fig3}d) follows a quasi-linear trajectory of 280 Hz/s, with a total frequency excursion of 17 kHz over 1 minute. However, it should be emphasized that the frequency variations of this spiral resonator, and more generally of all integrated devices, are neither linear nor are they monotonic over long time scales, due to the many parameters that can influence the waveguide's refractive index. As a consequence, our ISCL's drift is not predictable, and thus we report our fractional frequency instabilities in Fig.~\ref{fig:fig3}c using the raw frequency data itself without drift subtraction. 

\begin{figure}[t b !]
\includegraphics[width = 0.95 \columnwidth]{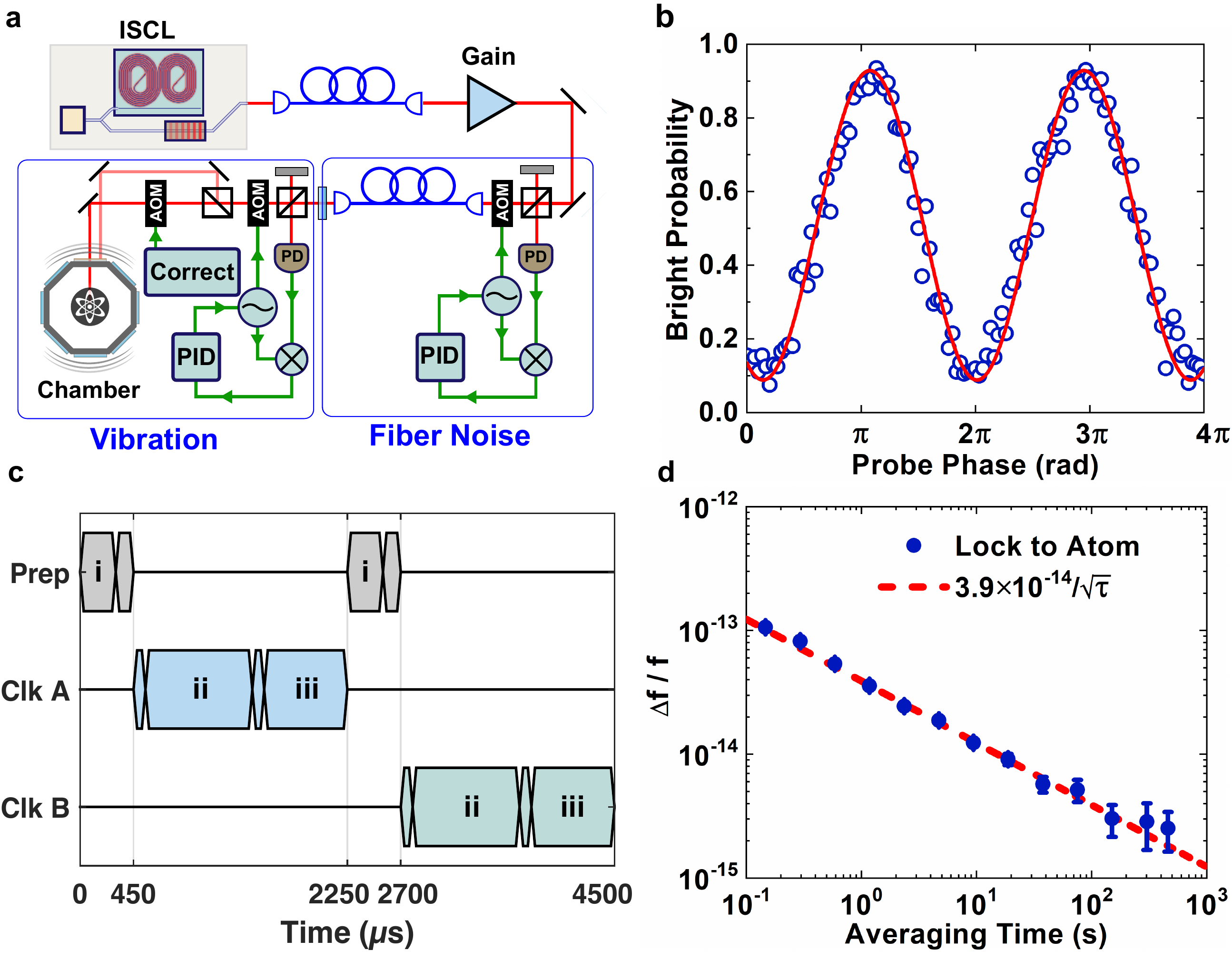}
\caption{
    \textbf{ISCL interrogation of an optical clock.}
    \textbf{a}, Schematic of laser beam path to the clock chamber. The chip resonator light passes through several fibers and amplification stages, as well as fiber noise and vibration cancellation systems. An AOM applies a final frequency offset used for
    spectroscopy and actuation of the atomic lock correction. 
    \textbf{b}, Measured Ramsey fringes on the clock transition.  Two $\frac{\pi}{2}$ pulses are applied around a $\tau=0.5\unit{ms}$ interrogation period.  The population in the lower clock state is measured as a function of the phase of the second pulse.
    \textbf{c}, Timing diagram for interleaved clock cycles. Blocks represent \textbf{(i)}, Doppler cooling ($300$~\mus) always followed by optical pumping, \textbf{(ii)}, $1$~ms precession with $\frac{\pi}{2}$ pulses immediately before and after, and \textbf{(iii)}, state detection.
    \textbf{d}, Fractional frequency instability of the difference between interleaved clocks.  The measured instability is divided by $\sqrt{2}$ to estimate the error of a single clock, assuming even distribution of error in the correction signals.  Vertical bars indicate $1\sigma$ errors. \cite{Howe2000}.
}
\label{fig:fig4}
\end{figure}

We validate the laser's performance against an atomic system by locking the laser to the narrow-linewidth $5S_{1/2}\rightarrow 4D_{5/2}$ transition in a trapped {$^{88}$\sr} ion. The ion is trapped $50~\upmu$m from the surface of a microfabricated ion trapping chip featuring niobium electrodes on a sapphire substrate. The chip is maintained in a cryogenic apparatus as described previously~\cite{Bruzewicz2019}. Care is taken to minimize additional phase noise during transmission from the ISCL to the ion, as illustrated in Fig.~\ref{fig:fig4}a. We utilize custom-built electrical circuits to compensate phase fluctuations associated with propagation in an optical fiber, to compensate for apparatus vibrations~\cite{Ma1994}, and to stabilize the amplitude of the interrogation laser. We characterize the performance of these systems using a light source stabilized to a ULE reference cavity. Using this reference, we find a Ramsey coherence time of $4.5~$ms, placing an upper bound on the unwanted contribution of residual fluctuations of the ion's transition frequency at the $1/(2\pi\times 4.5~\text{ms}) = 35~$Hz level.

We lock the ISCL to the ion via a Ramsey sequence which consists of two $\pi/2$ pulses separated by a free evolution time of 1 ms. The relative phase between the two pulses is fixed at $\pi/2$, mapping differential phase between the atom and laser during the free evolution period to differential quantum state amplitudes in the ion. After Ramsey evolution, we measure the ion state and apply a frequency correction to the laser depending on the outcome of the ion measurement.
The overall frequency update each clock cycle is given by
\begin{equation}
\centering
\Delta f_{i+1}=\Delta f_i + \delta \! f m_i+ \beta I_i
\end{equation}
\begin{equation}
\centering
I_{i+1} = \gamma I_{i} + \delta \! f m_i,
\end{equation}
where $i$ is the iteration number, $\Delta f_i$ $(\Delta f_{i+1})$ is the current (next) applied frequency correction, and $m_i$ is the sign $(\pm 1)$ of the correction which depends on the measurement result. $\delta \! f = \alpha /2 \pi \tau$ denotes the frequency correction step and is 21.0 Hz for our operating conditions of 1 ms interrogation time ($\tau$) and an $\alpha$ gain parameter of 0.13. $\beta$ represents the coefficient of an integral term that increases locking range in the presence of the laser's long-timescale drifts, $\gamma$ is the memory of the integral and determines how quickly the past contributions decay, and $I_i$ $(I_{i+1})$ denotes the cumulative integral correction of the current (next) iteration step (the initial term $I_1 = 0$). 
We also implement a startup sequence which spectroscopically locates the clock transition and initiates the lock with millisecond latency (See Methods Sec.~\ref{sec:clockprotocols}). 
To quantify the performance of the atom stabilized light source, we perform a Ramsey sequence interleaved with correction cycles, giving rise to the data shown in Fig.~\ref{fig:fig4}b. A fringe contrast of $80\%$ is observed at $500~\upmu$s when interrogating the ion with the ISCL, with an overall $1/e$ coherence time of about $2~$ms. 
To determine the short-term stability of our trapped-ion clock, we perform a self-comparison by operating two independent and interleaved feedback stabilization loops, each operated on the same hardware and ion but without knowledge of the other's atom-stabilized frequency, as shown in the timing diagram in Fig.~\ref{fig:fig4}c. This self-comparison measurement is insensitive to drifts of the ion's transition frequency itself, such as those from magnetic field fluctuations or ion micromotion, which would be common-mode to these interleaved clocks. Nevertheless, this self-comparison reliably captures the dynamics associated with the atom-laser feedback system~\cite{NicholsonClockComparison2012,NicholsonSystematicClock2015}. We obtain the fractional frequency instability shown in Fig.~\ref{fig:fig4}d, which averages down as $3.9 \times 10^{-14}/\sqrt{\tau}$.  

Our experimental results agree with a simulation of the laser lock to the ion that uses the measured laser noise and incorporates ion projection noise and the Dick effect due to finite clock dead time, as shown in Fig.~\ref{fig:fig3}c. The simulation uses an atom interrogation time of 1 ms and assumes a dead time between interrogation cycles of 3 ms. This configuration closely corresponds to our experimental conditions, due to our interleaved self-comparison. The simulated fractional frequency instability of the ISCL locked to the atom reaches $4.0 \times 10^{-14}$ at 1 second, closely agreeing with our experimental results. Under normal operating conditions of the clock and without the use of interleaved measurements, the dead time would be reduced to $\sim$1 ms, which we simulate would improve the clock instability to $2.1 \times 10^{-14}$ at 1 second, assuming no significant drifts currently common-mode to the interleaved clocks.

In the future, the use of on-chip arrays of multiple ions could improve the SNR of the ion measurement and thus allow correspondingly higher stability while also potentially fully eliminating dead time via sequential interrogation of multiple subarrays \cite{biedermann2013zero}.  
Another important step will be to operate in a non-interleaved manner and compare against a separate system or reference clock to validate the long-term stability of the clock. 
Finally, the use of a room-temperature vacuum system for the ion trap offers the potential for reduced acoustic noise and enhanced portability.


Our use of an ISCL for the interrogation of a strontium ion clock demonstrates that on-chip reference cavities are viable as a means to produce coherent light for the most advanced and demanding applications in atomic physics. Previously, it was shown that a fiber SBS laser could similarly replace the ULE cavity-stabilized lasers employed in conventional optical clocks \cite {Loh2020}---but the ability to integrate the entirety of the ISCL on chip has far greater implications for realizing true system portability. Our ISCL demonstration here advances the state-of-the-art in the frequency instability of integrated light sources to $7.5 \times 10^{-14}$, while our development of new protocols for performing spectroscopy and atom locking enable operation of the optical clock despite laser drift. We achieve a short-term clock instability of $3.9 \times 10^{-14}/\sqrt{\tau}$ from a time-interleaved clock measurement, demonstrating the potential for portable optical clocks to one day replace extant microwave clocks outside the laboratory.

\emph{Note - } During preparation of this manuscript we became aware of related work \cite{chauhan2024trapped}. 

\section{Methods}

\subsection{Spiral resonator design}

We design the resonator by maximizing the total path length achievable within the allotted 26~mm $\times$ 32~mm area defined by our reticle. To achieve dense packing, a spiral geometry is selected, where in-going and out-going spirals are interleaved, and connect to one another near the center at a point of inflection (Extended Data Fig.~1). In order to ensure first and second order differentiability at all points, we specify the entire structure in a piecewise analytic manner, while also applying strict constraints to waveguide spacing and minimum bend radius. The individual sections of the spiral are then stitched together to form a complete resonator ensuring no discontinuities in the light path. These requirements, however, do not preclude the possibility for abrupt changes in the directionality of the light, which we eliminate by enforcing first and second order differentiability along the waveguide path.  Our approach guarantees smoothness within each piece, but requires careful matching of derivatives across joints of the composite function. The relevant components are summarized further in Table~1. We also respect a maximum of $0.15~\text{mm}^{-2}$ for the first derivative of the curvature.

\renewcommand{\arraystretch}{1.5}
\begin{table}
\centering

\begin{tabular}{| c | c | l |}
\hline
Component              & Color  & Description \\
\hline
Circles & blue & Tightest curves closest to the inflection points. \\
\hline
Cubics  & red & Segments parameterizable as ($x,x^3$) link the circles. \\
\hline
Racetracks  &  \begin{tabular}{@{}c@{}}purple\\light blue\end{tabular} & \begin{tabular}{@{}l@{}}These are specified in a polar coordinate system by fitting \vspace{-2mm} \\a cosine expansion to a piecewise racetrack shape. \vspace{-2mm}\\ A small winding term is added, and additional loops are \vspace{-2mm}\\computed analytically with vector addition techniques \vspace{-2mm}\\to be precisely shifted by the required trace spacing.\end{tabular}\\
\hline
Interconnects  & green & Additional cubic segments connect the two racetracks.\\
\hline
Bus Waveguide  & black & 
\begin{tabular}{@{}l@{}}An In-coupling structure meets reticle edges at $8^\circ$ angles,\vspace{-2mm}\\ makes a close pass to the spiral resonator for in-coupling,\vspace{-2mm}\\ and uses cubic elements to respect differentiability.\end{tabular}\\
\hline
\end{tabular}
\caption{
\label{tab:components}
Various components of the piecewise analytic function describing the spiral resonator shape and bus waveguide. The color column gives the color of the component in Extended Data Fig.~1. 
}
\end{table}

The straight bus waveguide which allows light to couple to the resonator was patterned with a gap to the spiral of 1.23 $\upmu$m. This gap was determined by integrating the coupling of the bus waveguide to the spiral and targeting a coupling that when averaged over the entire spiral would constitute a loss of 0.1 dB/m. The bus waveguide width was set at 5.25 $\upmu$m to ensure single mode operation at 1348 nm.

\subsection{Spiral resonator fabrication}

The resonators were fabricated in MIT LL’s Microelectronics Laboratory. Fabrication starts with 200 mm wafers with 15 µm thick thermal oxide as supplied by the wafer vendor. Wafers undergo standard cleaning including piranha and SC1/SC2 (standard clean) chemistries and are then deposited with 40 nm of LPCVD stoichiometric Si$_3$N$_4$ in a tube furnace at 800°C. Following deposition, the nitride film is annealed in a nitrogen ambient at 1100°C for 3 hours. We then pattern the resonator structure with a 300 nm thick resist and 65 nm bottom anti-reflective coating (BARC) layer, exposing the full 26 x 32 mm reticle size to maximize the area available for a maximal length resonator. The full thickness waveguide structure is etched using a LAM TCP 9400 etch system with a SF$_6$ etch chemistry, and the resist is stripped in an O$_2$ plasma ash system. Scanning electron microscope (SEM) metrology is performed on a 500 nm wide linear structure to verify critical dimensions are within 15 nm of the design width ($\pm$ 15 nm 3$\sigma$). Step height measurements confirm that the full thickness of the nitride layer was etched with $<$10 nm of over-etch into the underlying oxide layer.

After the waveguide etch is complete the wafers are once again cleaned and loaded into the furnace for a series of TEOS SiO$_2$ based waveguide cladding layers. The 4 $\upmu$m thick cladding is deposited by four successive depositions of 1 $\upmu$m thick LPCVD TEOS layers at 700°C. After each 1 $\upmu$m layer is deposited, the film stack is annealed in a N$_2$ ambient at 1100°C for 3 hours, except the final oxide layer which is annealed at same temperature for 6 hours. The incremental deposition and anneal of the TEOS reduces the incidence of stress induced cracking in the cladding layer.

\subsection{Resonator Q and optical loss measurements}

We measure the spiral resonator's Q by sweeping a $\sim$100 kHz linewidth seed laser over the resonance lineshape of the three transverse modes that are confined by the spiral's waveguide geometry, and measuring the transmitted power with a photodetector. Due to the close frequency spacing of the spiral resonator's modes, a single narrowband scan covers all three transverse modes of the structure. The frequency axis of the scan is calibrated by performing a second scan with an intentionally applied modulation that generates sidebands of a known frequency spacing relative to the center laser line. For our spiral, we chose the modulation frequency to be between 3-5 MHz. A sweep over the spiral resonances once again yields the three main tranverse mode resonances along with smaller resonances that correspond to the sidebands crossing the modes, spaced apart from the carrier by the chosen modulation frequency.
\newline
\indent The intrinsic Q and waveguide loss are extracted from the measured resonance linewidth and extinction ratio for each of the transverse modes. These parameters are related through
\begin{equation}
\centering
\Gamma=\frac{1/\tau_e-1/\tau_0}{1/\tau_e+1/\tau_0}
\end{equation}
\begin{equation}
\centering
\frac{1}{\tau}=\frac{1}{\tau_e} + \frac{1}{\tau_0}
\end{equation}

\noindent where $\Gamma$ is the fractional field transmission amplitude past the resonator and $1/\tau$, $1/\tau_e$, and $1/\tau_0$ are the total, extrinsic, and intrinsic field decay rates of the resonator. In terms of measurement parameters, $\lvert\Gamma\rvert^2$ is the measured fractional transmitted power when operated at resonance, and $\tau$ is related to the measured resonance linewidth $(\Delta f)$ through $1/\tau=\pi\Delta f$. Solving this system of equations for the intrinsic waveguide loss rate yields
\begin{equation}
\centering
\frac{1}{\tau_0} = \frac{\pi \, \Delta f \, (1-\Gamma)}{2}
\end{equation}

\noindent Finally, after putting the loss rate in optical power units, we derive waveguide losses of 0.15 dB/m, 0.17 dB/m, and 0.24 dB/m from our measured resonance linewidths for the TE$_{00}$, TE$_{10}$, and TE$_{20}$ modes, respectively.

\subsection{ISCL operation}

The spiral resonator is operated with 10 mW of optical power at the fiber input to its bus waveguide. We measure optical losses at the input and output facets of the resonator chip to be 3.9 dB each, which results in 4.1 mW of power that enters the bus waveguide before being delivered to the input of the spiral resonator. The PDH modulation and demodulation is performed at 28 MHz, with the servo applying its feedback to the laser current to keep the seed laser locked on the spiral resonance. 
Changes in optical power delivered to the resonator that occur as a result of this feedback drive, as well as fluctuations in the power of the laser in general, are suppressed by the intensity servo applied before entering the spiral chip. The spiral rests in a copper enclosure having an area of 3$''$ $\times$ 2.25$''$ and a height of 1$''$. This copper enclosure is temperature controlled with a thermoelectric cooler and a thermistor at 23.5 °C. 15 mW (50$\%$) of the seed laser power is diverted to an SOA for boosting the level of the 1348 nm light prior to frequency doubling. We obtain 107.5~$\upmu$W of light at 674 nm, which is subsequently sent through a PM fiber optic cable to the trapped ion lab for further amplification and atom interrogation. 

\subsection{ISCL optimization}

We choose the optical power that couples into the spiral to optimize performance, requiring a tradeoff between several effects that influence the frequency stability of the laser. Extended Data Fig.~2a shows the measured frequency noise of the ISCL at 1348 nm for a portion of the spectrum between 100 Hz and 10 kHz offset frequency. The power that couples onto the spiral chip is varied from 1.1 mW to 5.1 mW via the current supplied to an SOA, and the results are compared to determine an optimal operating point. The plot shows that the baseline frequency noise remains relatively constant with optical power up until input power of 5.1 mW, at which point power-induced temperature fluctuations become significant, but noise at harmonics of the power line decreases with increasing optical power. This decrease of the power line noise at offset frequencies of 60 Hz, 180 Hz, and 300 Hz is better visualized in Extended data Fig.~2b, where the noise at power line harmonics is directly graphed against the on-chip optical power. The improvement of power line noise applies to all frequencies compared and is observed to be approximately an order of magnitude across the full range of optical powers. One possible explanation for this improvement is that the amplitude of the error signal increases with higher optical power relative to the contribution of line noise, thus increasing the signal-to-noise ratio of the lock.
\newline
\indent The effects of power line noise and power-induced temperature fluctuations combine to degrade the stability of the ISCL across a variety of timescales. Extended Data Fig.~2c depicts the fractional frequency noise compared across the same conditions of on-chip optical powers previously used, and for averaging times between 1 ms and 1 s. At low optical powers, the presence of large power line noise greatly increases $\Delta f / f$ below 30 ms. As the optical power increases, the stability improves until 4.1 mW of power is reached, where the stability degrades slightly between 3 ms and 20 ms. At this point, the power-induced thermal fluctuations become the limiting factor, as is especially noticeable in the measurements at 5.1 mW. As the observed noise at 1 ms (our Ramsey interrogation time) is lowest for an operating power of 4.1 mW, and this operating power also reaches the lowest overall fractional frequency noise at an averaging time of 30 ms (Extended Data Fig.~2d), we choose this input power for operating the optical clock with the ISCL.

\subsection{Optical clock simulation}

Our calculations of the expected clock operation are performed by comparing the measured time series of the ISCL's frequency against a simulated atomic reference, assumed to be perfectly stable in frequency in comparison to the laser. The time series of the laser is frequency doubled for operation at 674 nm, and its intrinsic sampling at 1 ms intervals corresponds exactly to the target interrogation time for clock operation. The laser frequency is assumed to be initially centered on a chosen Ramsey fringe, and later measured values of the laser frequency are compared to this fringe to determine the sign of the frequency correction applied. The dead time is accounted for by skipping points in the measured time series until the start of the next interrogation cycle (to a precision of 1 ms).

The Ramsey fringe is modelled by a sinusoid with a periodicity that is the inverse of the interrogation time. At each interrogation cycle, the position of the laser frequency relative to the center frequency of the fringe is tracked, and a probability is calculated and converted into a binomial random variable that determines the correction sign. A single step is applied to the laser's frequency along the direction of the chosen sign, with magnitude calculated according to the protocol outlined in Eqs. 1 and 2 of the main text, and these steps are accumulated as the simulation progresses. For our simulation, the next three points in the time series are skipped (corresponding to a dead time of 3 ms), and the process repeats until the end of the measured time series.

\subsection{Testing optical clock lock with drift}\label{sec:testingclockdrift}

We experimentally verify the benefits of adding an integral term to the atomic lock protocol by intentionally applying a 0.1 Hz sinusoidal modulation to the output of the interrogation laser while it interrogates the strontium ion. The amplitude of the sinusoid is gradually increased over time, and the ability of the applied corrections to track the added modulation is compared for the case with and without the integral term. For purposes of this experiment, we use a high-performance ULE-cavity stabilized laser operated directly at 674 nm, whose frequency drift is low ($\sim$60 mHz/s) compared to that of the applied modulation. 

Extended Data Fig.~3 demonstrates the clock corrections in response to the applied sinusoidal modulation, with the modulation varying from 0 to a maximum slope of 29 kHz/s. The use of a sinusoid enables testing not only of laser drifts in one specific direction, but also of regions where the drift turns around and accelerates in the opposite direction. With an interrogation time of $\tau=1$ ms and without integration applied ($\alpha= 0.2$), the atom lock holds up to a maximum slope of 3.1 kHz/s, but quickly deteriorates past this point. However, with integration introduced into the lock protocol ($\alpha= 0.13$, $\beta= 0.0035$, and $\gamma$=0.995), the atom lock remains intact until the slope reaches 14 kHz/s. As a point of reference, the maximum drift we have measured for the ISCL at 674 nm is $\sim$2 kHz/s. Although this drift is within the determined range where the laser can be locked to the atom without the use of integration, our tests in Extended Data Fig. 3 are based on sinusoidal applied modulations where the variations in laser drift are relatively gentle over time. To ensure robustness in the lock of our ISCL to the atom for a variety of drift conditions, we incorporate the integral term into our lock protocol.

\subsection{Startup sequence 
\label{sec:clockprotocols}}

In order to initiate the Ramsey clock on the correct fringe, we begin with a spectroscopy scan with a low probe power $500$~\mus ~pulse and $1~$kHz point spacing. We perform just $10$ trials per point and initiate the lock as soon as more than 4 of the trials result in ion excitation to the $D_{5/2}$ state, with power set low enough so this is only possible on resonance. By preloading this entire procedure on our FPGA we achieve latencies less than $10$~ms in shifting from the startup sequence to running the clock protocol. In practice this startup sequence almost always succeeds in detecting and locking to the ion on the first pass, although occasionally a second pass is required.



\section{Data availability}

The data sets that support this study are available on reasonable request.

\section{Code availability}

The code used for analysis and simulations are available on reasonable request.


\section{Acknowledgements}

 This work was sponsored by the Under Secretary of Defense for Research and Engineering under Air Force contract number FA8721-05-C-0002 and by the Defense Advanced Research Projects Agency (DARPA) under contract number FA8702-15-D-0001. The views, opinions and/or findings expressed are those of the author and should not be interpreted as representing the official views or policies of the Department of Defense or the U.S. Government. Distribution Statement “A” (Approved for Public Release, Distribution Unlimited)

\section{Contributions}

W.L. D.R., R.M., and D. G. conceived, designed and carried out the experiments with the ISCL.  D.R., R.M., W.L., and W.S. conceived, designed and carried out the experiments with the clock protocol. D.K. and A.S. fabricated the spiral resonators. C.B., R.T.M., and A.M. performed the resonator fiber attach. C.D.B. and D.R. developed the fiber and vibration noise cancellation for the clock system. All authors discussed the results and contributed to the manuscript.

\section{Competing interests}

The authors declare no competing financial interests.


\clearpage

\section{Extended data figures and tables}

\subsection{Extended Data Fig.~1 Spiral Resonator Component Sections}

\begin{figure}[H]
\centering
\includegraphics[width = 0.65 \columnwidth]{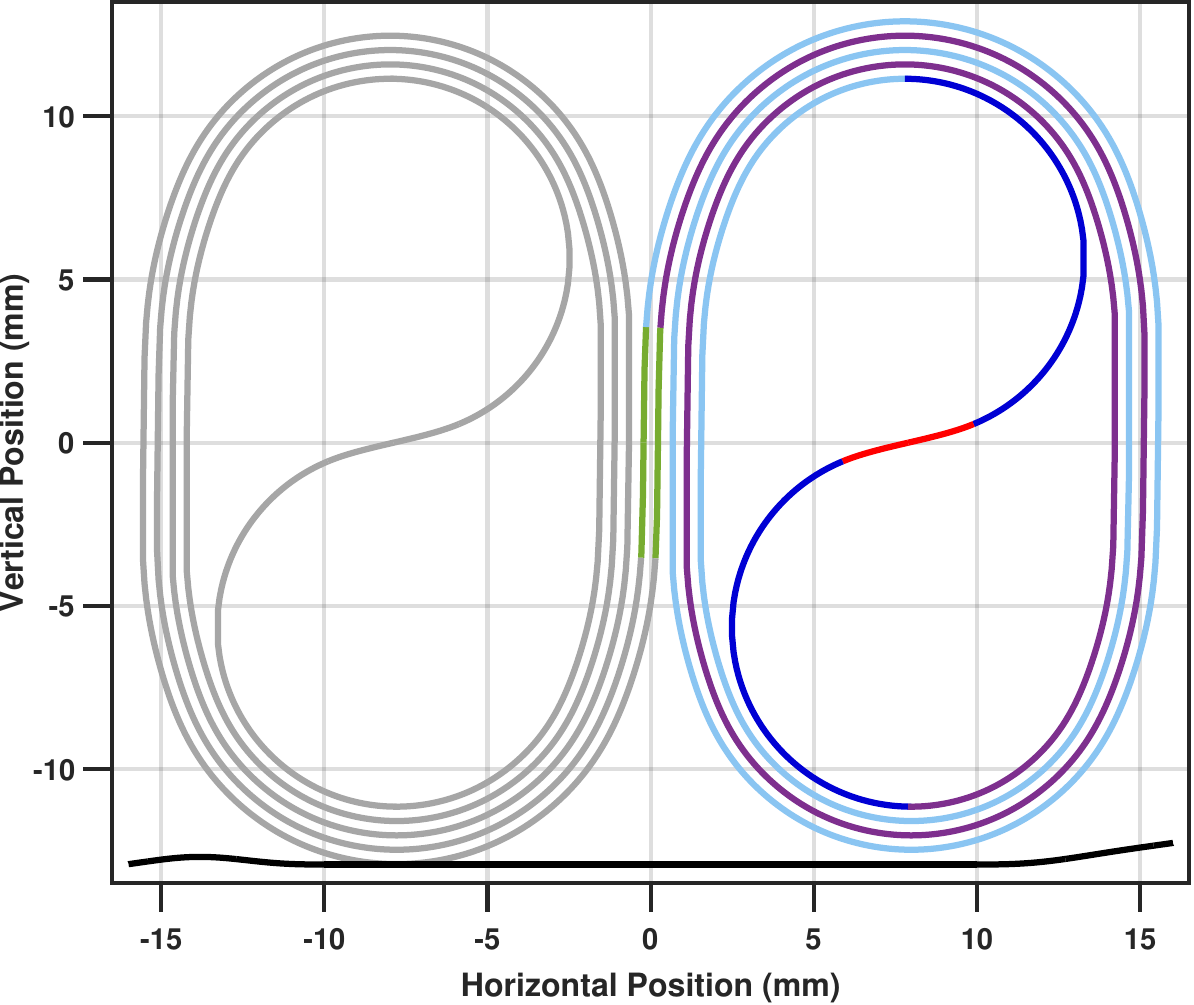}
\label{fig:extendeddata_fig1}
\end{figure}

\noindent
Plot of the spiral resonator with its different component sections highlighted. The number of revolutions is reduced from $25$ to $2$ for clarity. The spiral resonator comprises the following sections: pure circles (blue), central cubics (red), inward going racetrack (purple), outward going racetrack (light blue), a rotated clone of all the preceding (gray), racetrack interconnects (green), and bus waveguide (black).

\clearpage

\subsection{Extended Data Fig.~2 Optimization of ISCL Operation}

\begin{figure}[H]
\centering
\includegraphics[width = 0.95 \columnwidth]{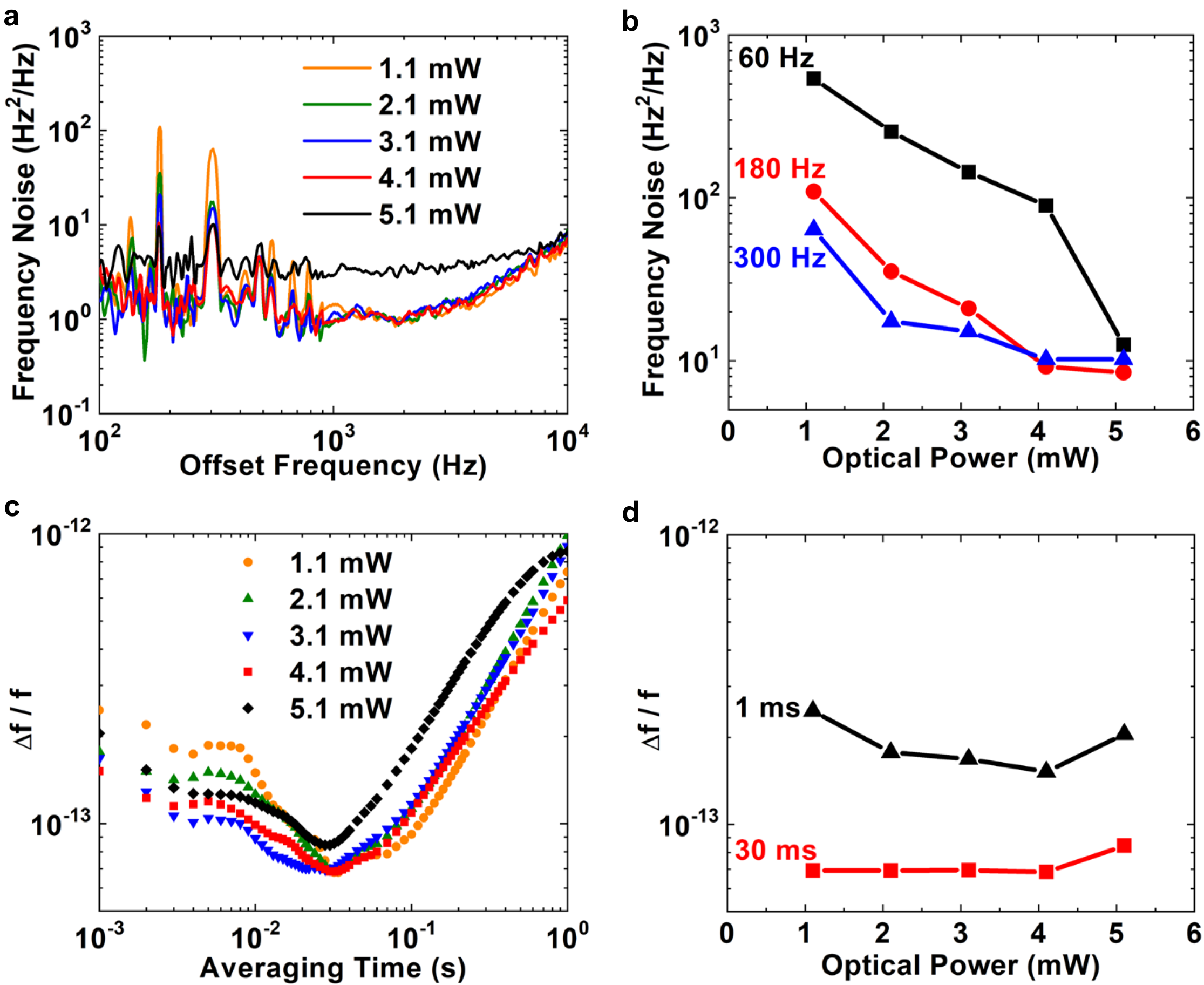}
\label{fig:methods_fig2}
\end{figure}

\noindent
\textbf{a}, Frequency noise spectra measured at a variety of chip-coupled optical powers. The line noise at 180 Hz and 300 Hz reduces with optical power. However, at 5.1 mW, the broadband noise level increases sharply.
\textbf{b}, Noise at the power line frequencies of 60 Hz (black solid squares), 180 Hz (red solid circles), and 300 Hz (blue solid triangles) plotted versus the on-chip optical power. The measured power line noise at all three frequencies decreases with increasing power.
\textbf{c}, Fractional frequency noise of the ISCL across the same range of optical powers previously tested. Changes in optical power lead to competing effects in the stability of the laser at both short and longer averaging times.
\textbf{d}, Fractional frequency noise versus optical power for averaging times of 1 ms (black solid triangles) and 30 ms (red solid squares). The optimum chip-coupled power is near 4.1 mW.

\clearpage

\subsection{Extended Data Fig.~3 Experimental Tests of Atom Locking With Integration}

\begin{figure}[H]
\centering
\includegraphics[width = 0.55 \columnwidth]{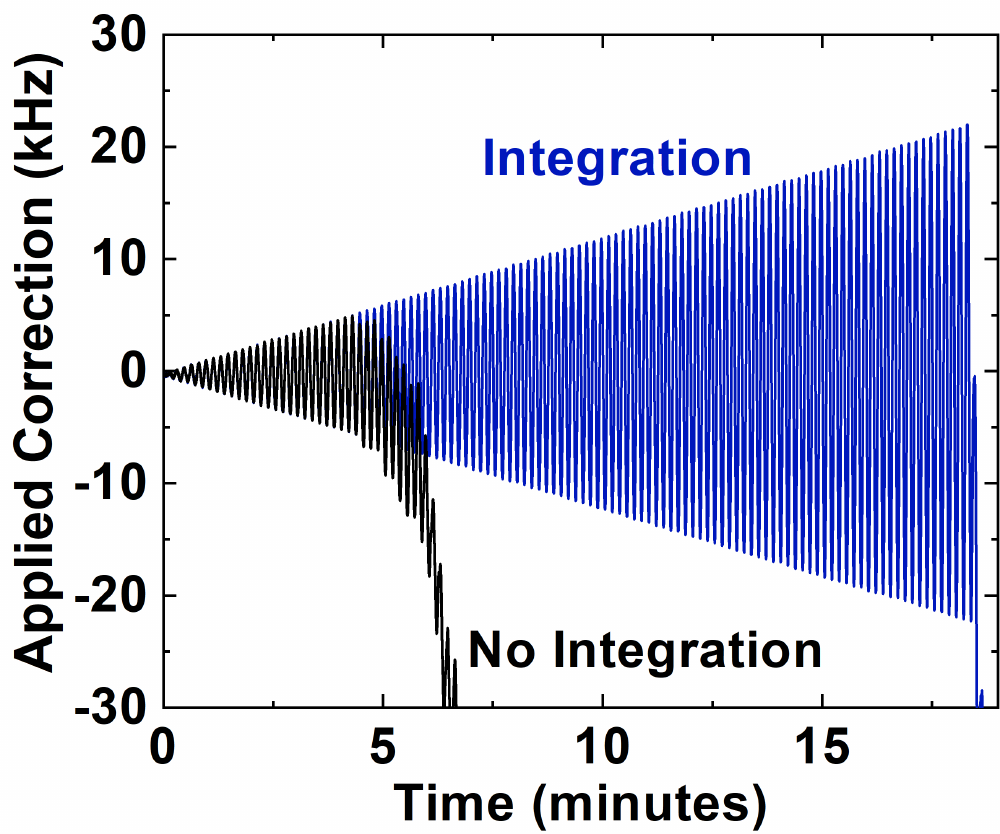}
\label{fig:methods_fig3}
\end{figure}

\noindent
Tests of the atom lock with $\tau=1$ ms and an intentionally applied 0.1 Hz sinusoidal modulation having an amplitude that increases over time. The trial where integration is included in the locking protocol (blue solid line, $\alpha= 0.13$, $\beta= 0.0035$, and $\gamma$=0.995) outperforms the case with no integration (black solid line, $\alpha= 0.2$).

\clearpage

\subsection{Extended Data Fig.~4 Atom Measurements Using the ISCL}

\begin{figure}[H]
\centering
\includegraphics[width = 0.95 \columnwidth]{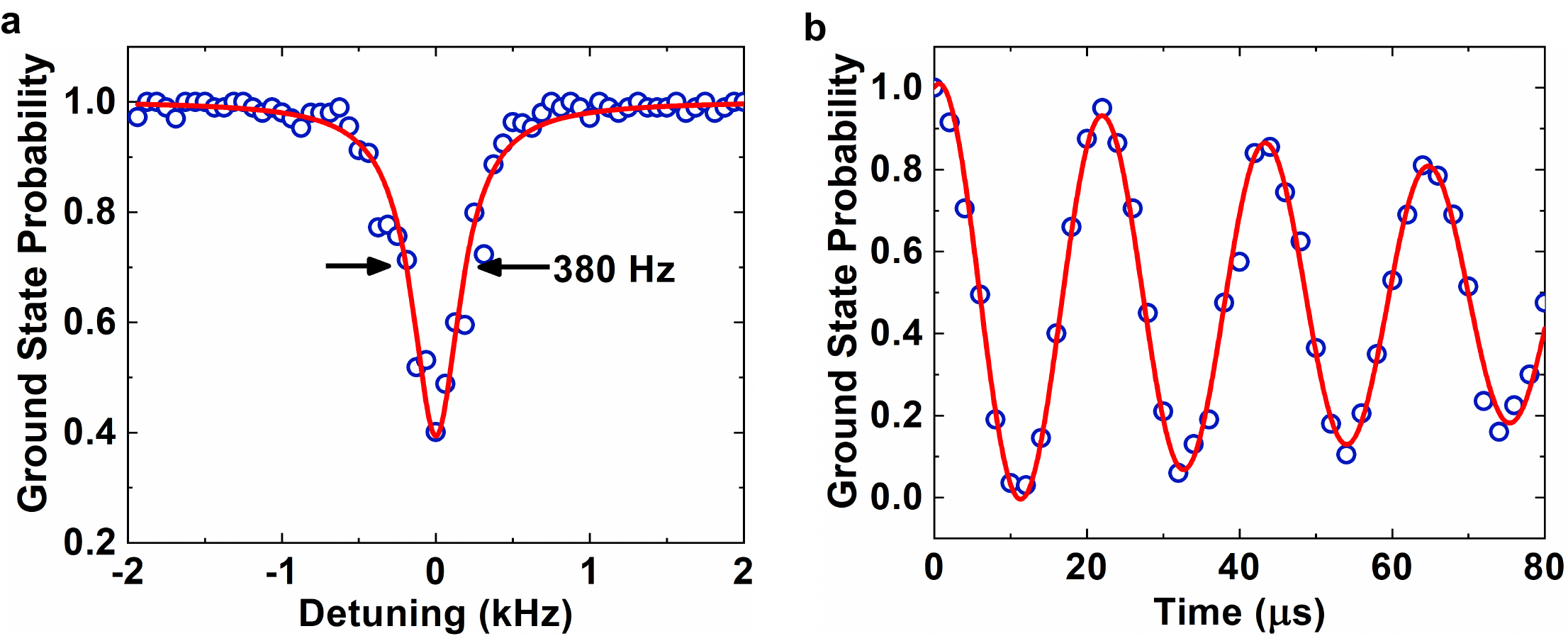}
\label{fig:methods_fig4}
\end{figure}

\noindent
\textbf{a}, Narrow spectroscopy performed with the atom and the ISCL. A $2.5$~ms probe pulse is used, increasing the deadtime of the clock and requiring greater performance. The observed linewidth of 380 Hz is Fourier limited by the interrogation time. 
The scan is performed with $400$ trials per point and was obtained in $\sim2$ minutes. 
\textbf{b}, Rabi spectroscopy performed with the ISCL. The fit demonstrates $\pi$-pulse time of $11$~\mus, with the observed decoherence due to the thermal state of the ion corresponding to mean axial mode occupation of $\bar{n} \approx 19.5$.

\clearpage


\bibliography{Chip_clock_biblio}

\end{document}